\documentstyle[psfig, twocolumn]{esapub}

\begin{document}

\setlength{\parindent}{0pt}
\setlength{\parskip}{ 10pt plus 1pt minus 1pt}
\setlength{\hoffset}{-1.5truecm}
\setlength{\textwidth}{ 17.1truecm }
\setlength{\columnsep}{1truecm }
\setlength{\columnseprule}{0pt}
\setlength{\headheight}{12pt}
\setlength{\headsep}{20pt}
\pagestyle{esapubheadings}
\title{\bf THE SOLAR INTERNAL ROTATION FROM GOLF SPLITTINGS}

\author{{\bf T.~Corbard$^1$,  M.P.~Di Mauro$^2$, T.~Sekii$^3$, 
and the GOLF team} \vspace{2mm} \\
$^1$Laboratoire G.-D Cassini, Observatoire de la C\^ote d'Azur, BP 4229, 
06304 Nice Cedex 4 \\
$^2$Istituto di Astronomia dell'Universit\`a di Catania, Viale A.Doria, 6 I-95125 Catania, Italy  \\
$^3$Institute of Astronomy, Madingley Rd., Cambridge, CB3 0HA, U.K. \\
}
\maketitle

\begin{abstract}

 The low degree splittings obtained from one year of GOLF data analysis 
are combined with the MDI medium-l 144-day splittings in order to infer
 the solar internal rotation as a function of the radius down to $0.2R_\odot$.
Several inverse methods are applied to the same data 
and the uncertainties on the solution as well as
 the resolution reachable are discussed.
The results are compared with the one obtained from the low degree splittings 
estimated from GONG network.    \vspace {5pt} \\


  Key~words: solar core rotation; inversion.

\end{abstract}

\section{INTRODUCTION}
The rotation of the solar core is an important  open question that
can be addressed by using spatial data from SOHO. In particular,
GOLF experiment  is dedicated to the observation of low-degree 
oscillations which sound the core. Here we use the GOLF frequency splittings
(\cite{golf})
 obtained from one year of observation beginning on April 11th 1996 together
with the MDI 144-day  splittings for degree up to $l=250$
(\cite{schou:mdi98}). For comparison we have also used the GONG splittings
of low-degree modes obtained (\cite{gong}) from 1 year of ground-based
observations (August 1995- August 1996).

\begin{figure}[h]
   \begin{center}
   \leavevmode
   \centerline{\psfig{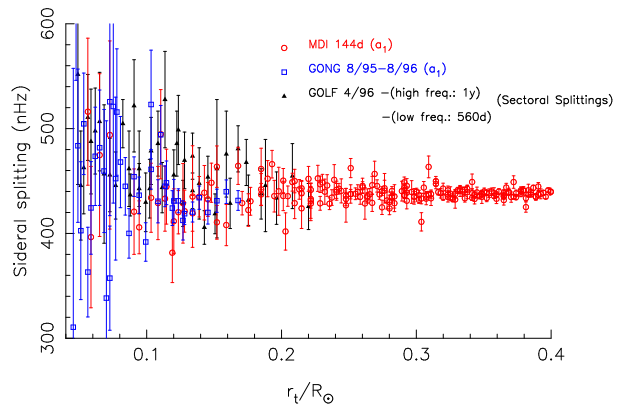}}
   \end{center}
 \caption{\em Sideral sectoral splittings (GOLF) and $a_1$ coefficients
(MDI and GONG) with their formal errors 
as a function of the turning points of the modes.}
 \label{fig:pt}
 \end{figure}

 Figure~\ref{fig:pt} shows the GOLF sectoral splittings  with their formal 
errors whereas  for MDI and GONG data the figure  shows the $a_1$ 
coefficients of the expansion of the splittings on \cite*{ritzwoller91}
polynomials. These two quantities (sectoral splittings and $a_1$ coefficients)
 may differ slightly in theory for $l>1$ because of the latitudinal
dependence of the solar rotation. In Section~\ref{sec:1D} we briefly
recall how the 2D inverse problem related to the internal rotation
can be reduced to a 1D problem for either the $a_1$ coefficients or the 
sectoral splittings and how in both cases 
the use of sectoral splittings may need some
a-priori assumptions on the rotation. 
 Then, Section~\ref{sec:methods} presents the different 
inverse methods used and, finally,
 we discuss the results obtained in Section~\ref{sec:results}


\section{FROM THE 2D TO THE 1D INVERSE PROBLEM}
\label{sec:1D}
For low rotation the frequency  of a mode of radial order
$n$ and degree $l$ is  splitted in $2l+1$ components of azimuthal order
$m=-l,..+l$ and the splittings $\Delta\nu_{nlm}\equiv{\nu_{nlm}-\nu_{nl0}\over m}$ are given by:

\begin{equation}\label{eq:split2D}
\Delta\nu_{nlm} \!=
\int_0^{1}\!\!\int_0^{R_\odot} \! K_{nlm}(r,\mu) 
\Omega(r,\mu) drd\mu,
\end{equation}
where $\Omega(r,\mu)$ is the unknown rotation rate versus depth $r$ and 
colatitude $\theta$ ($\mu=\cos\theta$) and $K_{nlm}(r,\mu)$
 the so-called rotational kernels calculated for each mode from oscillation eigenfunctions
of an equilibrium solar model.

In order to investigate the rotation below $0.4R_\odot$ where the latitudinal
dependence is particularly not well constrained because of the few 
 azimuthal orders provided by the low $l$ degree modes,
 one may want to simplify the problem and reduce it
to 1D problem in radius. As a matter of fact 2D inversions are usually not able
to peak averaging kernels in both radial and latitudinal direction below
$0.4R_\odot$ and two different 1D approximations of Equation~\ref{eq:split2D} 
are used instead.

\subsection{1D relation for sectoral splittings}\label{sec:split_1D}

One possibility in order to obtain a 1D integral relation is to search 
for the rotation rate at a given latitude
only. Since the most
constrained zone is the equatorial one, we try to investigate only the 
equatorial rotation rate. For this purpose we use the 
approximation
\begin{equation} \label{eq:2Dapprox}
K_{nlm}(r,\mu)\simeq K_{nl}(r)P_l^m(\mu)^2
\end{equation}
where $K_{nl}(r)$ are radial kernels (e.g. \cite{cuypers80}) and 
$P_l^m(\mu)$ are Legendre functions normalized such that 
$\int_{-1}^1 P_l^m(\mu)^2d\mu=1$,
 and which satisfy the following 
property  for $m=l$:
\begin{equation}\label{eq:Pll}
P_l^l(\mu)^2=C_l(1-\mu^2)^l\ \ \ 
C_l={(2l+1)!\over 2^{2l+1} (l!)^2 }
\end{equation}
This shows that, for high-degrees $l$, the  major contribution to 
sectoral splittings $\Delta\nu_{nll}$
comes from the equatorial rotation rate 
$\tilde\Omega_{eq}(r)=\Omega(r,\mu=1)$. This leads to the  1D
integral approximation:

\begin{equation}\label{eq:split_1D}
\Delta\nu_{nll}\simeq\int_0^{R_\odot} 
K_{nl}(r)\tilde\Omega_{eq}(r)dr
\end{equation}
Nevertheless this  approximation is valid only for high degrees $l$. For
lower degrees the sectoral splittings are sensitive not only to the equatorial
 rotation but also to the rotation rate in a
large angular domain  
around the equator. The extent of this 
 domain and the influence 
of this approximation on the estimation are discussed in 
\cite*{corbard:capodimonte}. 

\subsection{1D relation for $a_1$ coefficients}

Some  experiments like GONG and MDI produce 
a small number of 
the so-called a-coefficients of splittings expansions on a 
set of orthogonal polynomials (\cite{ritzwoller91}).
Assuming that the relation between individual splittings and these 
coefficients is linear, an equation similar to Equation~\ref{eq:split2D} 
can be 
established by computing the
 appropriate kernels $K_{nlj}^a(r,\mu)$
 related to each 
$a^{nl}_j$-coefficients for odd indices $j$ (see e.g. \cite{pijpers97}).
Furthermore it has been shown by \cite*{ritzwoller91} that  the expansion 
of the splittings in orthogonal polynomials corresponds to an expansion
of $\Omega(r,\mu)$ such that:
\begin{equation}\label{eq:expan}
\Omega(r,\mu)=\tilde\Omega_{1}(r)+\sum_{j=3,5...}\tilde
\Omega_{j}(r){dP_{j}(\mu)\over d\mu},
\end{equation}
where $P_j$ are the Legendre polynomials. This forms the so called 1.5D
problem where each $a_j$-coefficient is related to the expansion function of 
the same index through a 1D integral. Therefore the first term of the 
expansion Equation~\ref{eq:expan} which do not depend on the latitude
can be related to the $a_1$-coefficients through:

\begin{equation}\label{eq:a1_1D}
a^{nl}_1=\int_0^{R_\odot}K_{nl}(r)\tilde\Omega_1(r)dr
\end{equation}

The radial kernel $K_{nl}(r)$ is the same as in Equation~\ref{eq:2Dapprox}
but, 
from Equation~\ref{eq:expan}, the function $\tilde\Omega_1(r)$  
obtained by inverting $a_1$ coefficients corresponds to the searched 
rotation rate only where the rotation do not depend on the latitude. Otherwise
it corresponds to some  average over latitudes that can  be estimated by 
looking at the corresponding 2D averaging kernel 
(cf. Section~\ref{sec:howto} and Figure~\ref{fig:noy2D}).

\section{INVERSE METHODS}
\label{sec:methods}
We have used two kinds of inverse methods for solving the 1D integral equations.
\begin{enumerate}
\item A Regularized Least-Squares (RLS) method with Tikhonov regularization
(see e.g. \cite{corbard:capodimonte}). This is a global method which
gives a solution at all depths which fits the data at the best in the least 
square sense. This is a linear method and then the value of the rotation obtained at any radius $r_0$ is a linear combination of the data:

\begin{equation}
\hat\Omega_{\left|{eq\atop 1}\right.}\!(r_0)=\sum_{nl}C_{nl}(r_0)\left|{\Delta\nu_{nll}\atop a_1^{nl}}\right.
\end{equation}  

By replacing in Equation~\ref{eq:split_1D} or \ref{eq:a1_1D} we obtain:
\begin{equation}\label{eq:avk1D}
\hat\Omega_{\left|{eq\atop 1}\right.}\!(r_0)=\int_0^{R_\odot}\left(\sum_{nl} 
C_{nl}(r_0)K_{nl}(r)\right)\tilde\Omega_{\left|{eq\atop 1}\right.}\!(r)dr
\end{equation}
The function in parenthesis is called 1D averaging kernel at $r_0$, 
$\kappa(r,r_0)$. The result obtained at $r_0$ will be easier to interpret as 
a local average of the rotation when this kernel is well-peaked and without strong oscillatory behaviour.

\item Two `local'  methods which search directly the coefficients $C_{nl}(r_0)$
which are able to peak the averaging kernel near $r_0$. The two methods
differ essentially in the way  to localize the averaging kernel.
 The  SOLA (Subtractive Optimally Localized Average)
method (\cite{pijpers92}) fits the averaging kernel to a Gaussian 
function of given width whereas the MOLA method (Multiplicative OLA) 
(\cite{backus70}) simply gives high weights in the minimization process
to the part of the averaging kernel which are far from the target radius.    
In both case we use a regularizing parameter in order to establish a balance 
between the resolution and the error magnification reached at the target 
$r_0$.

\end{enumerate}

\section{HOW TO USE $l=2,3$ SECTORAL SPLITTINGS?}\label{sec:howto}

As already quoted, GOLF data for $l=2,3$ are not $a_1$ coefficients 
but sectoral splittings. Therefore one may want to use 
the Equation~\ref{eq:split_1D} in order to infer the equatorial 
rotation rate. There are two difficulties with this approach:
  \begin{figure}[h]
   \begin{center}
   \leavevmode
   \centerline{\psfig{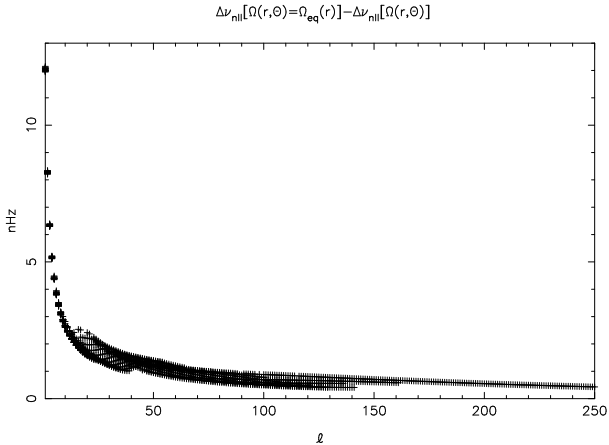}}
   \end{center}
 \caption{\em Difference in sectoral splittings computed from Equation~1 
and from the 1D approximation Equation~6 for a given 2D rotation profile 
resulting from a 2D RLS inversion of MDI data. The
     difference is around $12$nHz for $l=1$, $8$nHz for $l=2$, 
$6$nHz for $l=3$.
This is due to the
     fact that for low $l$, $P_l^l(\mu)$ extend far from the equator 
and therefore low-degree sectoral
     modes are more sensitive to the latitudinal dependency of the rotation
 in the convection zone  than other sectoral modes.}
 \label{fig:correc}
 \end{figure}

\begin{enumerate}

\item As already mentioned in Section~\ref{sec:split_1D}, the relation 
Equation~\ref{eq:split_1D} is not valid for low-degree $l$ and therefore
may not be suited for the determination of the core rotation. 
This may be corrected by assuming that the latitudinal dependence of 
the rotation rate (i.e. $\tilde\Omega_j(r)$ $j=3,5..$)
is known (taken from some previous 2D inversions for example). 
With this assumption, we can correct the observed sectoral splitting
prior to inversion by adding for each mode
 the difference between the sectoral 
splittings computed from Equation~\ref{eq:split2D} and from
Equation~\ref{eq:split_1D}. This difference is plotted on Figure~\ref{fig:correc}
as a function of the degree $l$.

\item MDI data do not provide the sectoral splittings for high $l$
      but only upon $18$ odd indexed $a$-coefficients. 
This number of coefficients 
is however high enough so that taking  their sum 
as sectoral splittings is a good approximation.  Nevertheless, the 
error on these sums (called  `truncated sectoral splittings' in the following)
is always higher than the error on $a_1$ alone.    
\end{enumerate}

Another possibility is to consider the sectoral splittings $l=2,3$
as $a_1$ coefficients at first approximation. In this approach
we do not need to correct the data because Equation~\ref{eq:a1_1D}
is valid even for low degrees. Nevertheless, we can also use other
dataset (MDI for example) or our 
knowledge of the latitudinal dependence of the rotation 
(taken from a previous 2D inversion for example) in order to estimate
$a_3$ for modes $l=2$ and $a_3$, $a_5$ for modes $l=3$.

With the same approximation as in Equation~\ref{eq:2Dapprox} we can write 
for the a-coefficients:
\begin{equation}
K^a_{nlj}(r,\mu)\simeq K_{nl}(r)Q_{lj}(\mu)
\end{equation}
so that, for both approaches,  the solution obtained at $r_0$ can ever
 be seen as an average of the rotation of the form:
\begin{equation}\label{eq:avk2D}
\hat\Omega(r_0)\!\!=\!\!\int\!\!\int\!\!\Bigl(\sum_{nl}C_{nl}(r_0) K_{nl}(r)W_l(\mu)\Bigr)\Omega(r,\mu)drd\mu
\end{equation}

The 2D averaging kernels (defined by the term in parentheses
 in Equation~\ref{eq:avk2D}) 
 can therefore be estimated from the 1D averaging kernels 
Equation~\ref{eq:avk1D}
 obtained at a target location $r_0$ by adding the angular part $W_l(\mu)$
of the 
2D rotational kernel  that corresponds to the data really inverted.

\begin{figure*}[t]
   \begin{center}
   \leavevmode
   \centerline{
                \hbox{\psfig{file=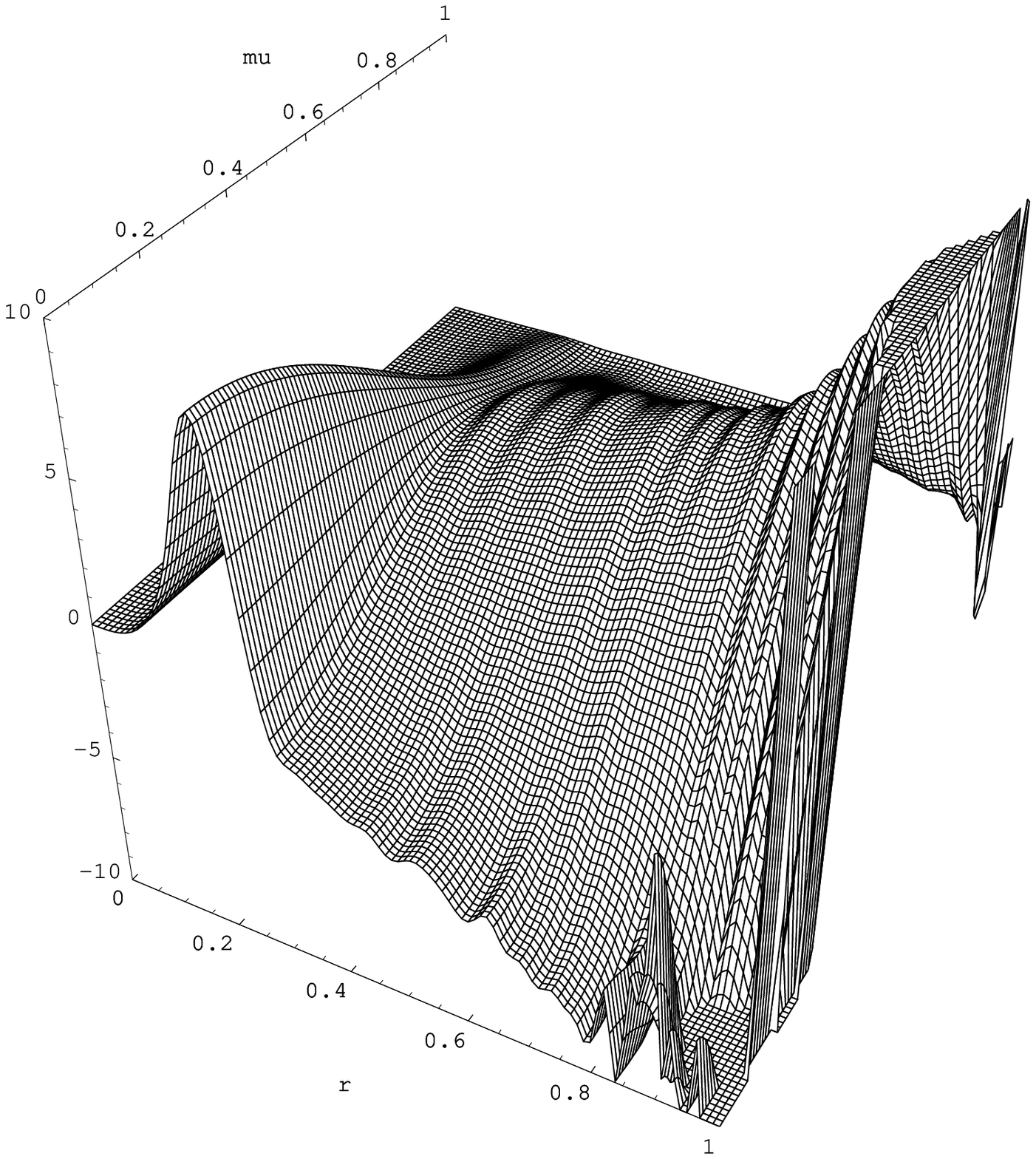,width=8.0cm,bbllx=100.pt,bblly=270.pt,bburx=597.562pt,bbury=770.pt,clip=t}
                     \psfig{file=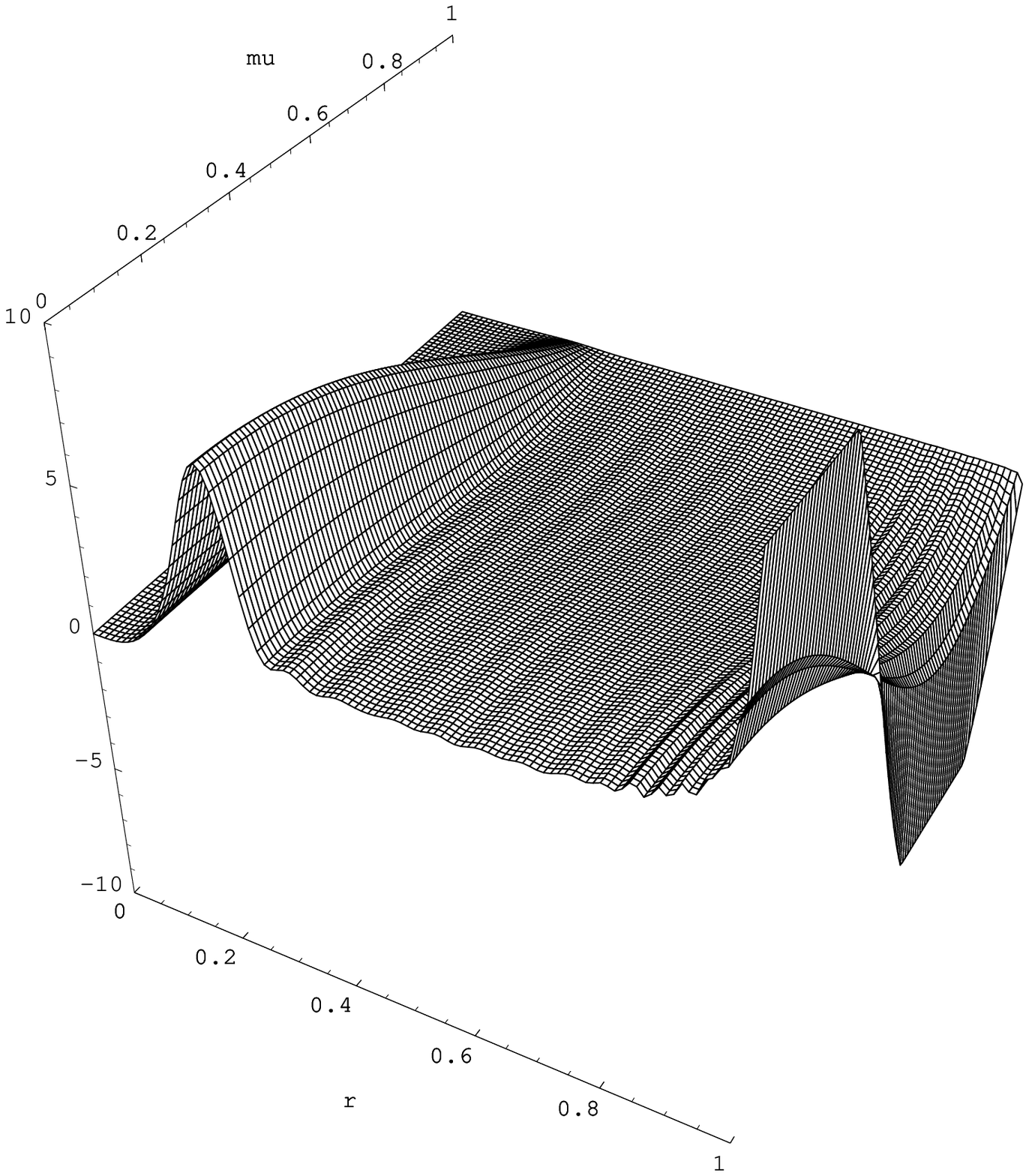,width=8.0cm,bbllx=100.pt,bblly=270.pt,bburx=597.562pt,bbury=770.pt,clip=t}
                      }
               }
   \end{center}
 \caption{\em 2D averaging kernels for 1D inversions computed at 
$r_0=0.2R_\odot$ (see Section~4). Left panel: inversion of GOLF sectoral 
splittings for $l\le3$ with MDI `truncated sectoral'(see text) splittings
for $l>3$. Right panel: inversion of GOLF sectoral 
splittings for $l\le3$ with  MDI $a_1$ coefficients. The peaks near the 
surface are truncated on the plot for clarity.}
 \label{fig:noy2D}
 \end{figure*}

Figure~\ref{fig:noy2D} shows on the left 
the 2D averaging kernels (defined by the term in parentheses
 in Equation~\ref{eq:avk2D}) obtained at $r_0=0.2R_\odot$
 by inverting
GOLF sectoral splittings together with MDI `truncated sectoral' splittings
 i.e.:

\begin{equation}
W_l(\mu)=\left\{
               \begin{array}{ll}
                 P_l^l(\mu)^2  &\mbox{for }l\le 3\\ 
                &\\
                \displaystyle \sum_{j=1}^{jmax} Q_{lj}(\mu)   &\mbox{for }l > 3 
               \end{array}	
         \right.
\end{equation}
whereas, on the right, it shows the 2D averaging kernel obtained at the same
target location by inverting MDI $a_1$ coefficients together with GOLF 
sectoral splittings for $l=1,2,3$ i.e.:
\begin{equation}\label{eq:ang_a1}
W_l(\mu)=\left\{
               \begin{array}{ll}
                 P_l^l(\mu)^2  &\mbox{for }l\le 3\\ 
                &\\
                 Q_{l1}(\mu)={3\over 2}(1-\mu^2)& \mbox{for }l > 3 
               \end{array}	
         \right.
\end{equation}

The two corresponding 1D averaging kernels (cf. Figures.~\ref{fig:molac} and 
\ref{fig:mola}) are the integral over latitude of 
these 2D kernels and are very similar: well localized near $0.2R_\odot$
 and without contributions near the surface. The angular part of the averaging
kernel for sectoral splittings strongly depend on the degree $l$ and this leads
to the oscillatory behaviour of the 2D kernel with very high peaks near 
the surface. From this plot it is clear that the interpretation of the result
obtained by inverting sectoral splittings is possible only if we have already 
a good knowledge of the latitudinal dependence of the rotation. Furthermore,
as already pointed out, this knowledge is needed in order to correct the low-degree
sectoral splittings for which the 1D approximation is not valid.
At the opposite,
for $a_1$ coefficients $Q_{l1}(\mu)$ is independent of $l$ 
(cf. Equation~\ref{eq:ang_a1}). Therefore
the surface oscillatory behaviour on the right panel of Figure~\ref{fig:noy2D}
comes only from the use of sectoral splittings for $l=2,3$. In this case,
the knowledge of the rotation profile is needed only near the surface in order
to interpret the result.

We can summarize the results of this study in few points:

\begin{enumerate}
\item
It is clear that it is better to use $a_1$ coefficients 
than sectoral splittings
when we have them.
\item In the case of GOLF data, we have access only to
 sectoral splittings for $l=2,3$. 
If we want to use them, it seems more reasonable
to try to correct  them by doing some assumptions on $a_3$ and $a_5$
for these modes
and to use the exact 1D integral 
rather than correcting all the  modes in order to use the approximated 1D
integral Equation~\ref{eq:split_1D}.
\item
In both approaches we can in principle obtain  a result easy to interpret
if the latitudinal
variation of the rotation is assumed to be known exactly.
 But by inverting 
$a_1$ for $l>3$ together with
sectoral splittings for $l<3$,
 we just have to make assumptions on the surface rotation. 

\item Several ways can be followed for correcting the results obtained 
by using sectoral splittings in $a_1$ inversions.
We can either 
\begin{itemize}
\item
take a guess rotation and integrating its surface part 
with the 2D kernel in order to correct the solution after the inversion or
\item take $a_3$ and $a_5$ from other datasets or 
\item calculate these coefficients from the guess rotation and subtract them 
from sectoral splittings before the inversion.
\end{itemize}
The best is probably to compare the effects of all these corrections
and to compare with the inversion of the corrected 
`truncated  sectoral' splittings. In any case assumptions are needed
on the latitudinal dependence of the rotation. As this dependence
can not be known exactly it  should be interesting to study in future works
how these assumptions increase the uncertainties on the solution.
\end{enumerate}  
The next Section shows some preliminary results obtained with these different
approaches for the use of the combined MDI and GOLF data.

\section{RESULTS AND DISCUSSIONS}
\label{sec:results}

\begin{figure}[h]
   \begin{center}
   \leavevmode
   \centerline{\psfig{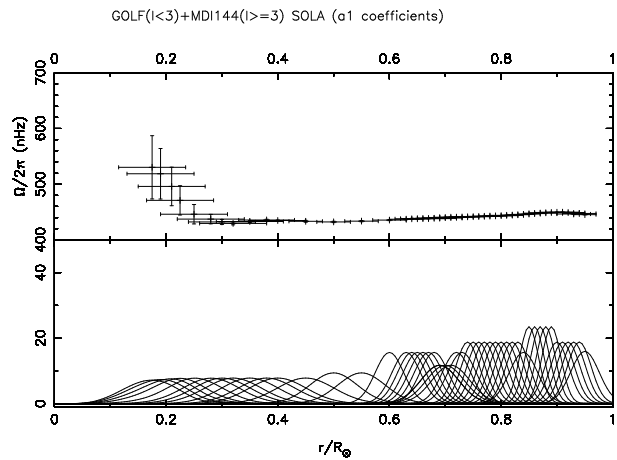}}
   \end{center}
 \caption{\em SOLA inversion of MDI $a_1$ coefficients and GOLF $l=1,2$
sectoral splittings. 
The corresponding 1D averaging kernels are shown below each
point of the solution. Vertical and horizontal error bars represent 
respectively the error and the resolution defined as the FWHM of the averaging kernel.  }
 \label{fig:sola}
 \end{figure}

\begin{figure}[t]
   \begin{center}
   \leavevmode
   \centerline{\psfig{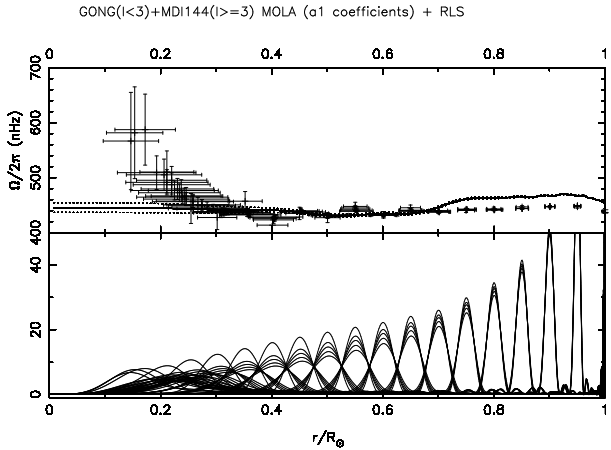}}
   \end{center}
 \caption{\em MOLA inversion of the same data as in Figure~4. Five regularizing parameters have been used at each target location. The full line shows an
 RLS inversion
of the corrected `truncated sectoral splittings' (see text) and the dotted 
line gives 1$sigma$ errors  on this solution. }
 \label{fig:mola}
 \end{figure}

\begin{figure}[!h]
   \begin{center}
   \leavevmode
   \centerline{\psfig{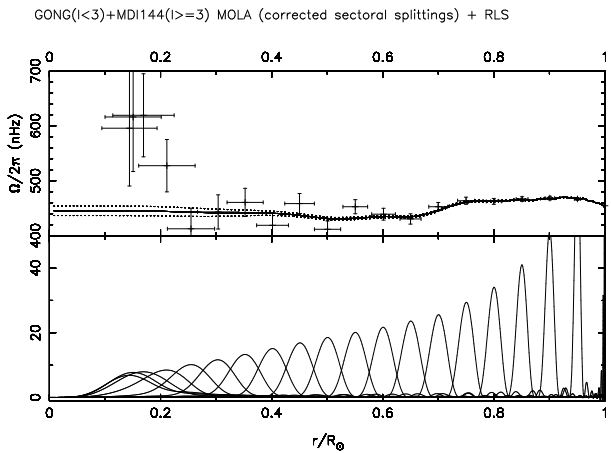}}
   \end{center}
 \caption{\em MOLA and RLS inversion of the corrected 
`truncated sectoral splittings'.
 Low regularization has been used in MOLA inversion. }
 \label{fig:molac}
 \end{figure}
\begin{figure}[!t]
   \begin{center}
   \leavevmode
   \centerline{\psfig{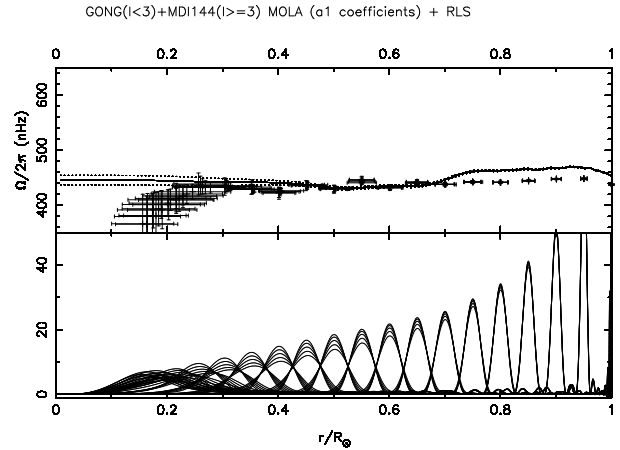}}
   \end{center}
 \caption{\em The same as Figure~5 but using GONG $a_1$ coefficients for l=1,2,3 instead of GOLF sectoral splittings.
}
 \label{fig:gong}
 \end{figure}

In this preliminary work, we have chosen, as a first step, to use
only $l=1,2$ GOLF sectoral splittings in order to reduce the difficulties 
discussed in the previous sections. Figures \ref{fig:sola} and \ref{fig:mola} 
show the results obtained by inverting the $a_1$ coefficients together
with $l=2$ sectoral splittings by using SOLA and MOLA inverse methods.   
In SOLA method the trade-off parameter is rescaled at each target location 
to obtain more localized averaging kernels. For MOLA method five trade-off
parameters have been used at each target location. In the case of low 
regularization MOLA averaging kernels have some small oscillatory parts.
The two methods fall in good agreement down to $0.20R_\odot$ where
the rotation rate increases up to $500\pm60$nHz. Below $0.15R_\odot$
both methods fail to peak kernels.
On Figure~\ref{fig:mola} we have shown a RLS solution obtained by inverting
sectoral splittings. As expected the two solutions ($\hat\Omega_{eq}$ and
 $\hat\Omega_{1}$) differ in the convection zone where the rotation rate 
vary with latitude. In the core,
 it is difficult to use RLS method because 
it is a global method and if one
try to obtain a well localized averaging kernel down to $0.2R_\odot$
 then we have to decrease
the regularization and the solution becomes very oscillating 
everywhere with big error bars. 
With an optimal L-curve choice of the regularizing parameter 
the solution is constant ($445nHz$ see Figure~\ref{fig:mola})
 below $0.4R_\odot$ but averaging kernels (not shown on the plot)
computed below this point 
are still localized near $0.4R_\odot$ 
so that, with this method,
 there is no conclusion on the core rotation in terms of weighted 
average of the true rotation. 
Nevertheless, we can use this solution and look at 
the residuals for each mode.
The global normalized $\chi^2$ of the inversion is 1.2.
 Now, if we look only at low-degree GOLF modes, 
the `partial normalized $\chi^2$' is around 0.5
showing that within error bars
GOLF data are in good agreement with a constant rotation below $0.4R_\odot$
 and that GOLF errors are probably not underestimated. Furthermore, looking
at the residuals for each individual splittings can help in the signal
analysis by pointing out some modes with high residuals that may be reanalyzed
in order to become more confident on the result.

Following the discussion of the previous sections we have also inverted
the sectoral (or `truncated sectoral') splittings corrected by using
the latitudinal dependence of the rotation found by a 2D RLS inversion of 
MDI data. The result is shown on Figure~\ref{fig:molac} in the case of low 
regularization. As expected the solution corresponds to the equatorial
rotation profile as found by the RLS method in the convection zone and the
 error bars increase compared to the use of $a_1$ coefficients alone.
The solution in the core is a little bit higher than found by $a_1$ inversion
but remains compatible within error bars showing also
an increasing rotation rate below $0.025R_\odot$. 

We have also tried to include GOLF $l=3$ sectoral splittings in our inversions.
This leads to a more  important increase of the rotation rate below 
$0.2R_\odot$ (around $700\pm100$ nHz). But in this case more work is needed
in order to become more confident in our result. In particular, in that
case, we have to  test the effects of the various corrections suggested
in Section~\ref{sec:howto} and to look at their influence 
on the estimation of the uncertainties on the core rotation.

Finally, we  have done the same analysis with GONG low-degree data
(\cite{gong}). In the case of GONG 
data we have $a_1$ coefficients so that, as quoted before,
 it may be more suited to carry an inversion of these
 coefficients using Equation~\ref{eq:a1_1D}
rather than using the `truncated sectoral' splittings.
In order to compare the result with the previous ones, the GONG $a_1$ 
coefficients for low-degrees have been used together 
with the MDI $a_1$ coefficients of higher degree modes.
 Figure~\ref{fig:gong} shows that
 these
data  tend to produce a slightly decreasing rotation rate below $0.3R_\odot$.
Therefore there is still a relatively important difference between 
the solutions obtained with the different low-degrees data. 
These differences are significant only if
 the error bars obtained on the solutions are not underestimated
and may be related  to the important dispersion of individual splittings 
measurements (cf. Figure~\ref{fig:pt}).  
Furthermore, we must notice that whereas MDI and
         GOLF splittings are for the same year of observations(5/96-5/97),
 the GONG data are for the year before (5/95-5/96) and therefore 
the results may not be directly compared. Therefore this result
 needs to be confirmed in future works and we have also to inverse  GONG $a_1$
coefficients for all the modes which should be a more self-consistent dataset. 

\section{FUTURE WORKS AND SOME QUESTIONS}
The problem of inferring the core physics remains
one of the most important still open  question  that can be addressed by 
 helioseismology. It is therefore very important, as a first step,
to be sure
that different `inverters' using different inverse methods 
can reach similar conclusions when they use the
 same datasets. This was the goal of this work in collaboration between
`inverters' within the GOLF team and we have shown that the different
approaches of the inversion are coherent. 
After this preliminary work several questions
need to be addressed, for example:
\begin{itemize}
 \item Can we obtain reliable results below $0.2R_\odot$ with actual 
datasets? 
\item How the results in the core are sensitive to the data used for 
medium $l$ splittings?
\item Can we explain the differences between the results obtained 
with different data by a systematic bias in some splitting measurements 
or by 
underestimated errors in a few number of individual splittings? 
\item 
Can these results (obtained with inverse methods) be confirmed by using
forward methods often used for the core 
rotation problem (\cite{charbonneau,gizon})?
\end{itemize}

\section*{ACKNOWLEDGMENTS}
We acknowledge  GOLF, 
SOI/MDI and GONG teams for allowing the use of data,  
and C. Rabello-Soares \& T. Appourchaux who provided us 
their analysis of 
GONG data for comparison.
SoHO is a project of international cooperation between ESA and NASA.
GONG is a project managed the NSO, a division of the NOAO, wich is operated by AURA, Inc. under a cooperative agreement with the NSF.
This work has been performed using the computing facilities provided 
by the program
``Simulations Interactives et Visualisation en Astronomie et M\'ecanique''
(SIVAM, OCA, Nice) and by the ``Institut du D\'eveloppement 
et des Ressources en Informatique Scientifique'' (IDRIS, Orsay). 
Thanks to the conference organizers for financial support.

\end{document}